\documentclass[fleqn,usenatbib]{mnras}

\usepackage{newtxtext,newtxmath}

\usepackage[T1]{fontenc}

\DeclareRobustCommand{\VAN}[3]{#2}
\let\VANthebibliography\thebibliography
\def\thebibliography{\DeclareRobustCommand{\VAN}[3]{##3}\VANthebibliography}

\usepackage{graphicx}	
\usepackage{amsmath}	

\newcommand{\gaia}{\textit{Gaia }}

\title[Binary fractions in GCs]{Fewer Companions in the Crowd: The Low Close Binary Fraction in Globular Clusters from \gaia RVS}

\author[Bashi \& Belokurov]{
Dolev Bashi,$^{1}$\thanks{E-mail: db975@cam.ac.uk} \&
Vasily Belokurov,$^{2}$\\
$^{1}$Astrophysics Group, Cavendish Laboratory, University of Cambridge, JJ Thomson Avenue, Cambridge CB3 0HE, UK\\
$^{2}$Institute of Astronomy, University of Cambridge, Madingley Road, Cambridge CB3 0HA, UK\\
}

\date{Accepted XXX. Received YYY; in original form ZZZ}

\pubyear{\the\year{}}

\begin{document}
\label{firstpage}
\pagerange{\pageref{firstpage}--\pageref{lastpage}}
\maketitle

\begin{abstract}
In dense environments like globular clusters (GCs), dynamical interactions can disrupt or harden close binaries, nonetheless, detailed comparisons with field binary fractions remain limited. Here, we present an analysis of the close binary fraction in a carefully selected sample of field stars and $10$ GCs using \gaia Radial Velocity Spectrometer (RVS) data, which is among the largest samples of GCs analysed using multi-epoch spectroscopy to date. By assessing the peak-to-peak variations of the sources' radial velocity (RV), we estimate the close binary fractions through a method that fits the distribution as the product of two Gaussian distributions. By applying the same RV-variability method to both cluster members and field stars, we ensure a homogeneous and inclusive comparison between the two environments. Despite matching stellar parameters between the field and GC samples, our findings confirm that GCs possess a significantly lower close binary fraction than field stars. Interestingly, we do not detect any clear trend of binary fraction with cluster metallicity – metal-rich and metal-poor GCs are uniformly binary-poor (within uncertainties). We discuss possible interpretations, including dynamical hardening in dense environments and the effects of common envelope evolution, which may lead to companion accretion or merger events. 
\end{abstract}




\section{Introduction}

Binary star systems are fundamental to the formation and dynamical evolution of globular clusters (GCs). In dense clusters, close gravitational encounters ionise and evaporate wide binaries. Mass segregation then brings the remaining tighter systems to the core, where subsequent stellar interactions harden them \citep{Heggie75,HurleyAarsethShara07, RoznerPerets24}. There, their interactions release energy that counteracts gravitational contraction and delays core collapse \citep{Gieles11}. Their presence offers insights into the initial conditions of cluster formation and the dynamical processes that shape these systems \citep{Hut92,Ivanova2005,Ivanova06}. Moreover, it is assumed that binaries are the progenitors of exotic objects such as blue straggler stars \citep[BSS;][]{Bailyn95}, millisecond pulsars, and X-ray binaries \citep{Clark75}, which are more common in GCs than elsewhere \citep{Carrasco-Varela25}. Dissolving clusters may also seed the `Aurora' \citep[the `in situ' Milky Way halo population;][]{BelokurovKravtsov22,Ardern-Arentsen25}, further
emphasising the long-term Galactic impact of binaries that originated in clusters. Thus, understanding the binary fraction in GCs is essential for refining dynamical models and comprehending stellar interactions within these systems.

Identifying and characterising binary systems in dense GCs, however, is observationally challenging. Traditional photometric methods rely on resolving a binary sequence on the colour-magnitude diagram (e.g. an excess of bright or colour-offset stars), which requires deep, high-resolution imaging and careful crowding corrections \citep[e.g.,][]{Milone12}. Eclipsing binary surveys can find some close binaries via periodic dips in brightness, but such systems are relatively rare and require long-term monitoring. The most direct approach—multi-epoch spectroscopy to detect radial velocity (RV) shifts — has historically been time-consuming and resource-intensive. For example, \cite{Sommariva09_M4} conducted an extensive campaign on the GC M4, collecting nearly $6000$ spectra with VLT/FLAMES to identify binary candidates. While spectroscopic observations provide valuable orbital parameters, they have been limited to only a few clusters due to the extensive telescope time required. Similarly, photometric surveys, for instance, HST-based studies \citep{Milone12, JiBregman15} and more recent JWST observations \citep{Milone25}, often must sacrifice field of view to achieve high resolution. Consequently, the binary populations of many globular clusters remain poorly constrained.

Decades of dedicated photometric and spectroscopic studies have already established that present-day binary fractions in Galactic GCs are markedly low, typically less than $ \sim 10\%$ for periods $ < 10^4$ days once selection effects are accounted for \citep[see e.g.,][]{Milone12, Lucatello15_BinaryGC, Sommariva09_M4, Kamann20_NGC3201, Wragg24_oCen, Horn25_47Tuc}. In contrast, a higher close binary incidence is found among comparably evolved field giants \citep[e.g.,][]{Moe19}. Numerical scattering experiments and cluster-scale simulations \citep{Heggie75, Ivanova2005} show that frequent three- and four-body encounters in dense cluster cores efficiently disrupt or harden primordial binaries, driving the global binary fraction downward and biasing the survivors toward hard, centrally concentrated pairs.

The advent of \textit{Gaia}’s Radial Velocity Spectrometer (RVS) has opened a new avenue for binary detection in GCs on a large scale. \gaia RVS provides repeated RV measurements for millions of stars across the sky \citep{Katz23}.
While \textit{Gaia}’s RVS provides valuable data across GCs, its observations are predominantly limited to brighter giant stars, often in the clusters’ outer regions. This limit poses a challenge in densely populated core areas. Specifically, the RVS faces constraints in crowded fields due to onboard processing limitations and window conflicts, which hinder accurate measurements in these regions \citep{Katz23, Weingrill23}. Consequently, these factors restrict the RVS’s effectiveness in the densest parts of GCs. On the other hand, the key advantage is the sheer scale and uniformity of the \gaia dataset: every star brighter than the RVS limit has dozens of RV measurements over the mission lifetime, allowing detection of RV variability indicative of binary motion. In particular, the \gaia DR3 catalogue provides for each star a robust estimate of the peak-to-peak RV variation -- RV$\text{pp}$ (see Section~\ref{sec:Sample} below for details). Stars with large RV$\text{pp}$ values are likely to be in close binaries, since orbital motion can induce significant RV variations, whereas single stars or long-period binaries show little variation.

Building on this approach, recent studies by \cite{Bashi24} and \cite{BashiTokovinin24} have demonstrated a powerful statistical method to infer binary fractions from \gaia RVS data. Rather than confirming individual binaries, their method models the distribution of RV semi-amplitudes (e.g. RV$_\text{pp}/2$) as a mixture of two populations – one for intrinsic single-star RV scatter and one for orbital motion of binaries – using a Gaussian Mixture Model (GMM). By fitting this mixture to the observed RV variability as a function of magnitude, one can deconvolve the contributions of binaries and singles and estimate the overall close binary fraction 
in the sample. This statistical technique overcomes many limitations of earlier binary surveys: it leverages the enormous sample size of \gaia and does not require resolving each binary’s orbit. Crucially, it allows for binary fraction measurements in numerous clusters in a homogeneous way. The \gaia DR3 RVS data set is unprecedented in its coverage, enabling us to derive binary fractions for an order of magnitude more clusters than was feasible with traditional multi-epoch spectroscopy \citep[e.g.][]{Sommariva09_M4, Lucatello15_BinaryGC, Giesers19_NGC3201,Wragg24_oCen,Horn25_47Tuc}.

In this work, we apply the RV variability method to \gaia DR3 RVS observations of 10 Galactic GCs, alongside a comparison field star sample, to robustly measure the close binary fraction in each environment. 
In Section~\ref{sec:Sample}, we describe the sample selection; Section~\ref{sec:Results} presents the close binary fractions in GCs and compares them to the field; Section~\ref{sec:discussion} discusses the implications of these findings.

\section{Globular Cluster members in \gaia RVS}
\label{sec:Sample}
We began with \gaia DR3 RVS sources having reliable RV measurements (\texttt{rv\_method\_used = 1}), i.e. bright stars ($G_{\rm RVS}<12$) and available peak-to-peak RV amplitudes, $\mathrm{RV_{pp}}$, (\texttt{rv\_amplitude\_robust}) values and required \texttt{parallax} > 0.
The \texttt{rv\_amplitude\_robust} field provides the total amplitude in the RV time series, defined as the difference between the maximum and minimum robust RV values after outlier removal. To identify these outliers, the \gaia team excluded valid transits with RV values outside the range Q1 - 3 × IQR and Q3 + 3 × IQR, where Q1 and Q3 are the 25th and 75th percentiles, and IQR = Q3 - Q1 (interquartile range). We focused on cool stars, i.e. $3900$ K < \texttt{rv\_template\_teff} < $6900$ K 
to avoid large RV errors in hot stars \citep{BlommeHotRV23} and at least $9$ independent RV observations (\texttt{rv\_visibility\_periods\_used} > 8) to ensure good phase coverage. 
As a final cleaning step, we cross-matched our sample with the \gaia RR-Lyrae catalogue \citep{Clementini23_RRLyrae} and removed potential RR-Lyrae candidates, which are more common in GCs and can exhibit large RV variations unrelated to binarity.
Using these cuts, we obtain $4,897,811$ sources.

To identify GC members, we used the catalogue from \cite{Vasiliev&Baumgardt21}, which leverages \gaia EDR3 astrometry to provide comprehensive data on Galactic GCs. This catalogue offers precise measurements of mean parallaxes and proper motions for $170$ clusters, enabling accurate membership determination. By cross-matching our data with this catalogue and adopting a membership probability threshold of $90\%$, we ensured the selection of genuine GC members for our analysis. This resulted in a final sample of $886$ sources distributed across $51$ different GCs within a distance limit of $d < 10$ kpc\footnote{This distance restriction was imposed to avoid a bias toward selecting only the brightest stars—such as those on the asymptotic giant branch—in more distant clusters, which would in turn skew the binary fraction estimates.}. Notably, $10$ of these clusters have more than $25$ RV-measured members; we list these clusters along with their distances ($d$), metallicities ([Fe/H]) and clusters' half-light radius ($r_{\mathrm{h}}$) as given in the catalogue of \cite{BaumgardtHilker18}\footnote{For a more recent version see also- 

\noindent \url{https://people.smp.uq.edu.au/HolgerBaumgardt/globular/parameter.html} } in Table~\ref{tab:GC}.

\begin{table}
	\centering
	\caption{Properties of the $10$ GCs in our Gaia RVS sample. Columns list the cluster name, number of member stars $N$ used, distance $d$, metallicity [Fe/H], half-light radius $r_{\mathrm{h}}$, and the median RV scatter radius $r_{\mathrm{RV}}$ (the median projected radius of the RV sample stars).}
	\label{tab:GC}
	\begin{tabular}{lccccc} 
\hline
Name & $N$& $d$ [Kpc] & [Fe/H] & $r_{\mathrm{h}}$ [pc] & $r_{\mathrm{RV}}$ [pc]\\
\hline
47Tuc & 155 & 4.52 & -0.72 & 4.52 & 12.66  \\
$\omega$Cen & 111 & 5.43 & -1.53 & 5.43 & 20.25  \\
M4 & 85 & 1.85 & -1.16 & 1.85 & 3.70  \\
NGC 6752 & 66 & 4.12 & -1.54 & 4.12 & 8.12  \\
NGC 4372 & 50 & 5.71 & -2.17 & 5.71 & 7.24  \\
NGC 3201 & 37 & 4.74 & -1.59 & 4.74 & 8.87  \\
M13 & 32 & 7.42 & -1.53 & 7.42 & 6.48  \\
NGC 6397 & 28 & 2.48 & -2.02 & 2.48 & 5.65  \\
M10 & 26 & 5.07 & -1.56 & 5.07 & 6.79  \\
NGC 4833 & 25 & 6.48 & -1.85 & 6.48 & 6.18  \\

		\hline
	\end{tabular}
\end{table}

\begin{figure*}
   \centering
\includegraphics[width=18cm] {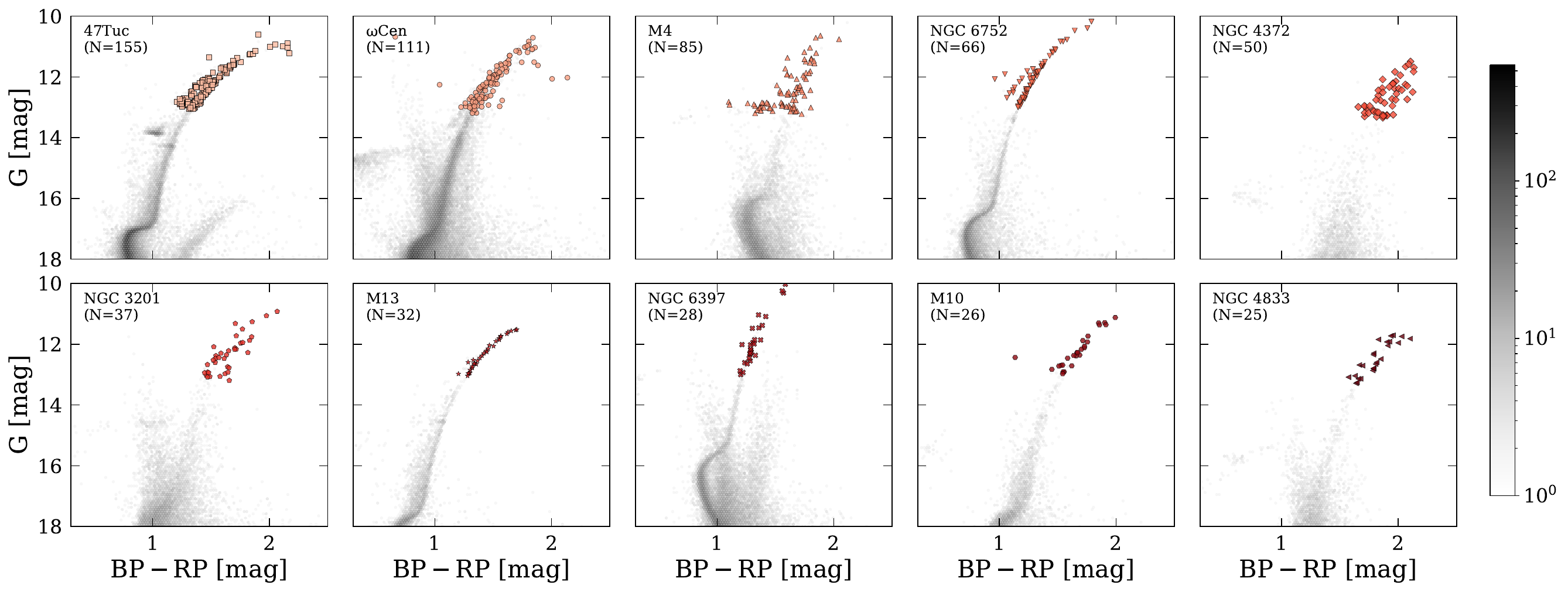}
   \caption{CMDs of the 10 GCs analysed in this work.  Grey-scale density maps show all high-probability members from the \citet{Vasiliev&Baumgardt21} catalogue, while coloured symbols highlight the subset with \gaia DR3 RVS information. The number of RVS stars in each cluster is given in parentheses.} 
              \label{fig:GC_bp_Rp_g}%
    \end{figure*}

We show in Fig.~\ref{fig:GC_bp_Rp_g} a Colour–magnitude diagram (CMD) of the stars contained in each of the $10$ clusters with \gaia RVS information. The background grey-scale colour marks a density plot of all members of the cluster as listed in \cite{Vasiliev&Baumgardt21}. As can be clearly seen, the RVS stars of each cluster populate the bright end of each CMD, composed mainly of red-giant-branch and horizontal-branch stars.    

\begin{figure}
   \centering
   \includegraphics[width=0.45\textwidth] {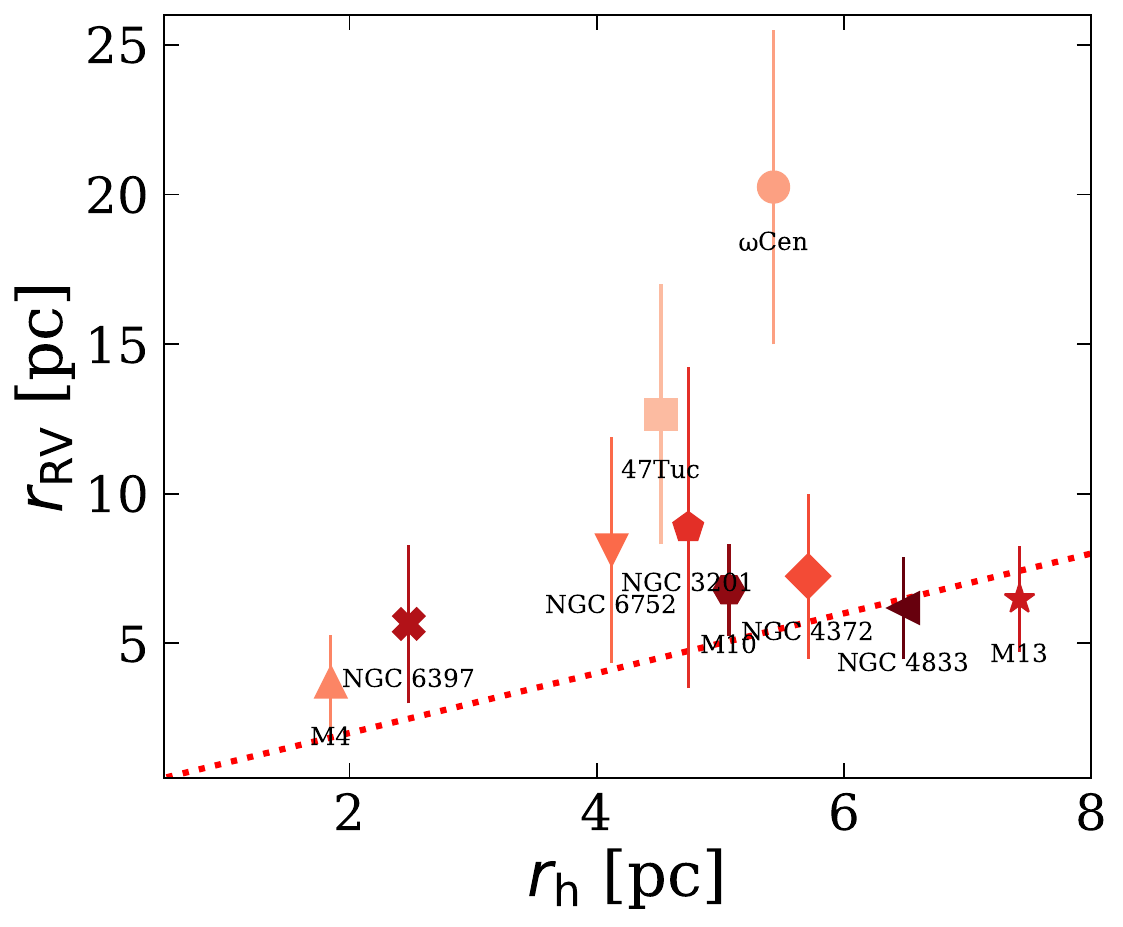}
   \caption{Median RV scatter radius ($r_{\mathrm{RV}}$) and MAD (y-error) for each globular cluster as a function of the cluster half-light radius ($r_{\mathrm{h}}$). The red dashed line marks a 1:1 relative ratio.}
              \label{fig:rh_rrv}%
    \end{figure}

To explore the scatter of sources with RV measurements compared to the overall cluster properties, we show in Fig.~\ref{fig:rh_rrv} the median RV scatter radius of each individual cluster, $r_{\mathrm{RV}}$ with its median absolute deviation (MAD) as the y-error, as a function of $r_{\mathrm{h}}$, where we define $r_{\mathrm{RV}}=d \times \sqrt{(x^2+y^2)}$ with $x$, $y$ are the projected angular positions of the sources relative to the cluster centre. We find the two properties consistent with each other, given the red dashed line marking a 1:1 ratio with one clear outlier, which is $\omega$Cen. This may be due to the limited coverage within the $\omega$Cen's dense field core and the \gaia DR3 astrometric crowding
limitations \citep{Weingrill23, Vernekar25} leading to \textit{Gaia}’s RVS primarily sampling stars in the outskirts rather than the denser central region of the cluster.

\section{Close binary fractions in Globular Clusters vs. the field}
\label{sec:Results}
To disentangle the contributions of single and binary stars in the \gaia RVS dataset, we adopt the same modelling approach as in \cite{BashiTokovinin24}. Briefly, we model the distribution of sources in the $G_{\mathrm{RVS}}$–$\log\left(\mathrm{RV_{pp}}/2\right)$ plane as a mixture of two Gaussian components - one representing single stars and the other representing binaries. For full details of the model formulation, priors, fitting procedure as well as our completeness assessment and correction method, we refer the reader to Appendix~\ref{app:methods}.

\begin{figure*}
   \centering
\includegraphics[width=17cm] {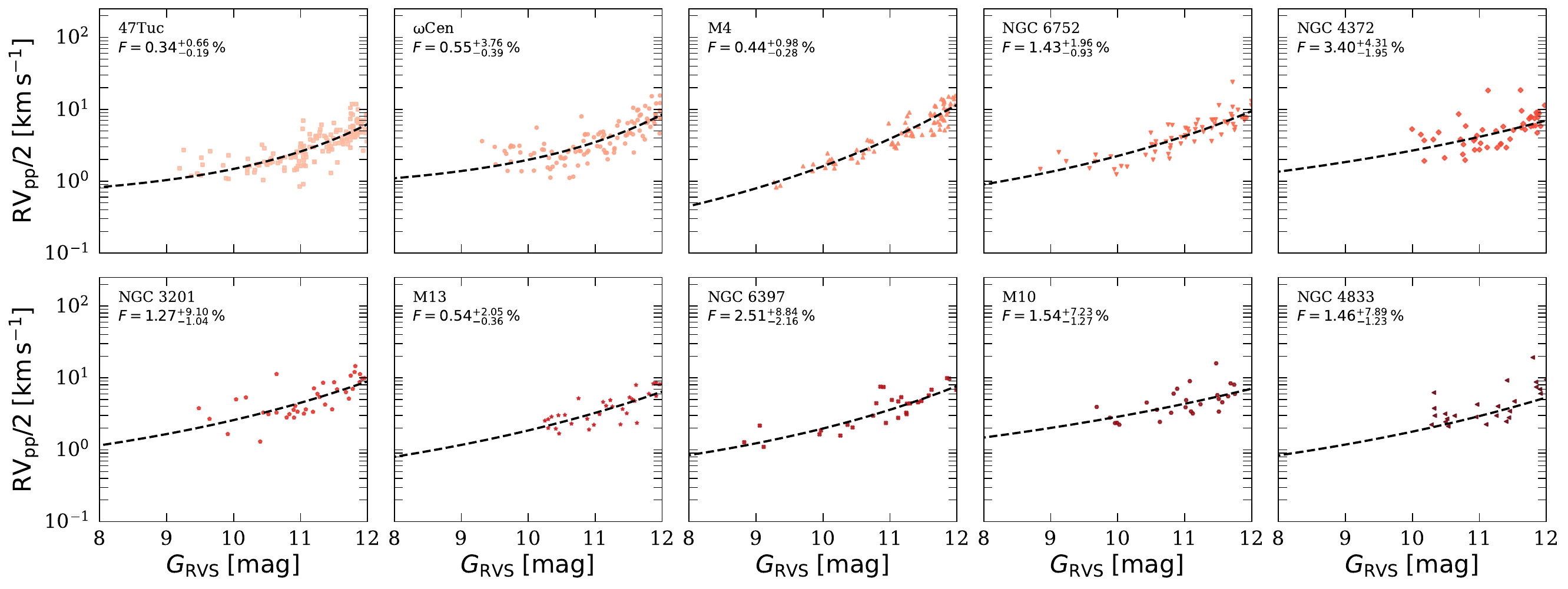}
   \caption{Mixture-model decomposition of \textit{Gaia}-RVS variability for the 10 GCs.
Each panel shows $\mathrm{RV_{pp}}/2$ versus $G_{\rm RVS}$ for all cluster members. The solid red curve traces the posterior median locus of the single-star Gaussian component recovered by the Bayesian mixture fit. The inset in each panel prints the raw close binary fraction $F$ returned by the fit (median and 16–84 \% interval; no completeness correction applied here). The scarcity of high-amplitude outliers relative to the narrow single-star locus translates into the very low binary fractions reported for most clusters; the completeness-corrected values are listed in Table~\ref{tab:GC_frac}.}
              \label{fig:GC_RVfits_grid}%
    \end{figure*}

Using our methodology, we show in  Fig.~\ref{fig:GC_RVfits_grid} the best-fit posterior values of the main trend of single stars $\mathrm{RV_{pp}}/2$ with $G_{\rm RVS}$ as well as a printout of the derived binary fractions. In most GCs, the number of binaries found is consistent with zero, yielding very low binary fractions.

Table~\ref{tab:GC_frac} presents the close binary fractions (median and $16\% - 84\%$ uncertainties) using our completeness correction method aimed to be complete up to $10^4$ days, along with those reported in previous works for the various GCs. Overall, the estimated binary fractions we find are consistent with those of earlier studies, while the variations can be attributed to several factors, including differences in detection methods, sample selections, and the specific stellar populations analysed. 

\begin{table}
	\centering
	\caption{Comparison of binary fractions in GCs from this work and previous studies. Note that the values reported here represent the close binary fraction ($\bar{F}$) in systems with orbital period $P\leq10^4$ days following Appendix~\ref{app:methods} completeness correction}, while literature values typically account for somewhat different range of binary population fractions ($F^*$) detected by various methods (see Section~\ref{sec:discussion} for further discussion).
    
	\label{tab:GC_frac}
	\begin{tabular}{lccl} 
\hline
Name & $\bar{F}$  [\%] & $F^{*}$  [\%] & \\
 &  this work & previous works & \\
\hline
47Tuc & $0.49^{+0.95}_{-0.28}$ & $2.4 \pm 1.0$ & \cite{Horn25_47Tuc}\\
$\omega$Cen & $0.80^{+5.45}_{-0.56}$ & $2.1 \pm 0.4$& \cite{Wragg24_oCen}\\
M4 & $0.63^{+1.42}_{-0.41}$ & $3.0\pm 0.4$ &\cite{Sommariva09_M4}\\
NGC 6752 & $2.07^{+2.84}_{-1.35}$ & $\approx 4$&\cite{MoniBidin08_NGC6752}\\
NGC 4372 & $4.93^{+6.26}_{-2.82}$ & ---&---\\
NGC 3201 & $1.85^{+13.20}_{-1.50}$ & $6.75\pm 0.72$& \cite{Giesers19_NGC3201}\\
M13 & $0.79^{+2.98}_{-0.52}$ & $2.4\pm 0.6$ & \cite{Milone12}\\
NGC 6397 & $3.64^{+12.82}_{-3.13}$ & $2.8\pm5.2$ & \cite{Milone12}\\
M10 & $2.23^{+10.48}_{-1.85}$ & $7.4\pm0.6$ & \cite{Dalessandro11_M10}\\
NGC 4833 & $2.12^{+11.44}_{-1.79}$ &$4.0\pm0.6$ & \cite{Milone12}\\
		\hline
	\end{tabular}
	\vspace{0.5em}
\end{table}

\begin{figure}
   \centering
   \includegraphics[width=0.48\textwidth] {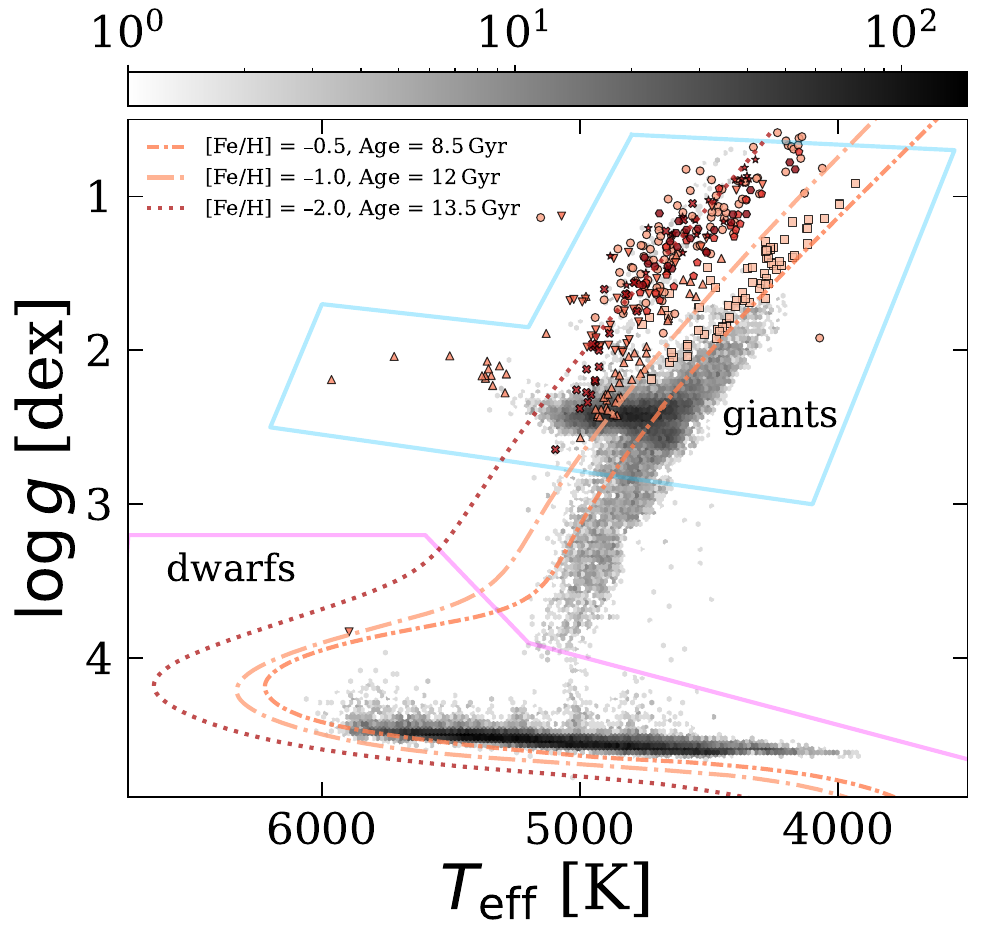}
   \caption{Effective temperature, \( T_{\mathrm{eff}}, \) as a function of the surface gravity, $\log g$, of the \textit{Gaia}-APOGEE field sample. The solid magenta and blue polygons delineate our selection of dwarf and giant sources, respectively. Overlaid are stellar GC members in our sample with available APOGEE stellar parameters. Dashed/dotted coloured curves show three representative MIST evolutionary tracks.} 
              \label{fig:kiel}%
    \end{figure}

To contextualise the GC results, we assembled a comparison sample of field stars cross-matched from \gaia DR3 and APOGEE DR17 \citep{APOGEE17}. We separated field stars into dwarfs and giants based on polygon regions on the Kiel diagram of surface gravity, $\log g$, as a function of effective temperature, $T_{\mathrm{eff}}$, shown in Fig.~\ref{fig:kiel}. We restricted the sample to $M_\ast \leq 1,M_\odot$ to avoid the higher binary fraction seen in more massive stars \citep{Raghavan10}. To do so, we used the extended catalogue of \cite{Stone-Martinez24APOGEEMass}, which lists the stellar mass of APOGEE sources.
We then applied quality cuts to the APOGEE data, removing stars with known issues such as poor spectra, very low S/N, or problematic RV solutions using the APOGEE \texttt{STARFLAG} and \texttt{ASPCAPFLAG} indicators; see \cite{Price-Whelan20}. To validate the polygon boundaries, Fig.~\ref{fig:kiel} also displays the available \gaia RVS GCs members as well as isochrones interpolation using the MESA \citep{MESA} Isochrones and Stellar Tracks (MIST) models \citep{ MIST0, MIST1}, which include three representative GCs metallicity and age values.

\begin{figure*}
   \centering
\includegraphics[width=16cm] {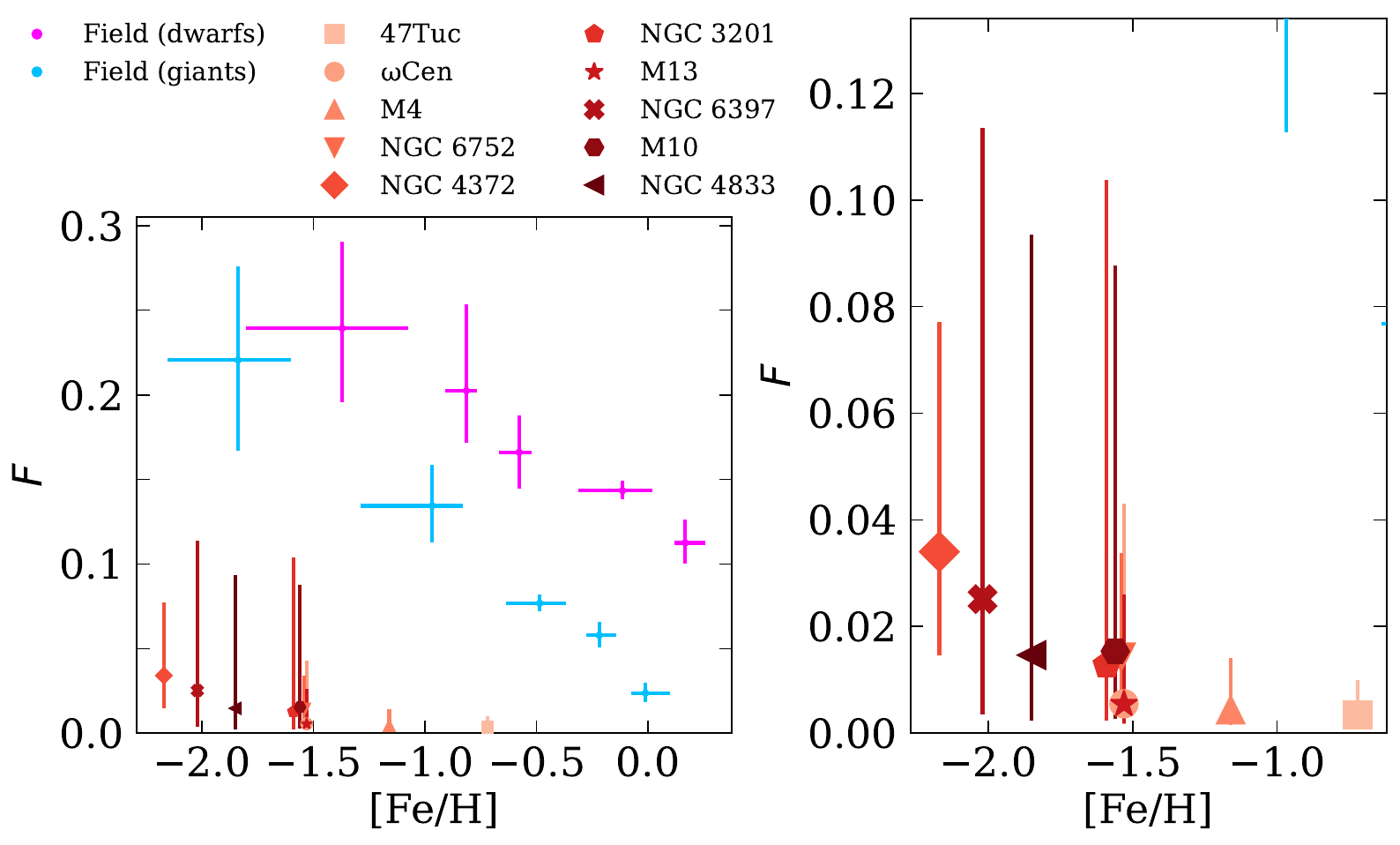}
   \caption{Close binary fractions as a function of metallicity for field stars and globular cluster (GC) members. Field star samples, comprising dwarfs (magenta crosses) and giants (blue crosses), were selected by cross-matching the \gaia sample with the APOGEE catalogue, providing metallicity information. Selection was based on the polygons marked in Fig.~\ref{fig:kiel}. Metallicity data for GC members were obtained from \citet{BaumgardtHilker18}. The right panel zooms in on the metal-poor region populated by GC members.} 
              \label{fig:feh_F}%
    \end{figure*}

Fig.~\ref{fig:feh_F} shows the close binary fractions as a function of [Fe/H], based on APOGEE measurements for field stars (both dwarfs and giants), together with the GC sample results. We choose not to correct for completeness the binary fractions displayed; consequently, the fractions derived in each metallicity bin should be considered 
complete mainly to short-period binaries, i.e., $P\leq 1000$ days, and intermediate mass-ratio $q=0.3-0.9$, and should be viewed as relative indicators rather than precise absolute frequencies. This distinction does not affect the robustness of the inter-bin or field-versus-GC comparisons that follow. In the field, we observe a clear anti-correlation between metallicity and binary fraction, with well-defined negative slopes of $-0.079$ for dwarfs and $-0.116$ for giants. These trends are consistent with earlier findings in the literature \citep[e.g.,][]{Moe19,Price-Whelan20}, as well as with our own analysis using the same methodology in \cite{Bashi24}. We also confirm that giants consistently show lower binary fractions than dwarfs across the metallicity range, which reflects evolutionary effects.

Against this backdrop, the GC sample stands out in stark contrast. Although GCs are composed predominantly of evolved stars and are therefore more directly comparable to the field giant population, their close binary fractions are markedly lower, typically just a few per cent or less, even at similar metallicities. Moreover, the GC data do not show a significant metallicity dependence within the current uncertainties where a linear fit yields a slope of $-0.021 \pm 0.025$, indicating no statistically significant trend. The insignificant slope contrasts with the clearly negative slopes seen in the field star samples. Nevertheless, we note that while the more metal-rich GCs, i.e. 47 Tuc, M4, exhibit binary fractions that are consistent with zero within uncertainties, the more metal-poor clusters tend to show mildly elevated values, although these offsets remain within the $1-\sigma$ error and are not formally significant.

\section{Discussion}
\label{sec:discussion}

By incorporating \gaia DR3 RVS data for cluster members, we achieve a homogeneous and statistically robust measurement of binary fractions across $10$ clusters, exceeding earlier spectroscopic studies that were limited to only a few clusters or small samples. This expanded scope allows us to confirm known trends and reveal new insights with much higher confidence. 

Our results confirm that close binary fractions in GCs are extremely low, consistent with earlier findings that binaries are rare in dense clusters \citep{Sommariva09_M4,Milone12, Kamann20_NGC3201, Horn25_47Tuc}. Only a few per cent (or less) of GC stars in our sample show the RV variability indicative of short period companions, which is much lower than the binary fraction in field populations. As most stars in our Gaia RVS GC sample are evolved giants, any close companion would likely have interacted with the primary during its expansion phase. As the star ascends the red giant branch, a close binary may undergo mass transfer or enter a common-envelope phase, frequently leading to coalescence or envelope ejection \citep{Paczynski76, Ivanova13}. This evolutionary channel naturally removes or merges short-period binaries before they can be observed. These merged systems may appear as single, more massive BSS, or overluminous giants \citep{Clark75}. The observed correlation between BSS frequency and present-day binary fraction across clusters supports this interpretation \citep{Sollima08, Rain24, Carrasco-Varela25}, suggesting that many close binaries have evolved into BSS or similar products. 

While our estimates agree with previous studies, differences between our work and earlier results likely arise from variations in detection methods, radial location selections within the cluster sample, and the evolutionary stage of the stars observed (i.e., dwarfs vs. giant stars). Mass segregation also plays a role: primordial binaries are expected to migrate toward the cluster core, where they may be more likely to survive or form via exchanges\footnote{Moreover, binary fraction in GCs may follow a `U-shaped' radial trend—low at intermediate radii due to core sinking, and higher in the outskirts where relaxation times exceed cluster ages. \citep[see, e.g.,][]{Dalessandro11_M10}.}. However, \text{Gaia}’s RVS measurements avoid the most crowded regions due to blending limitations \citep{Katz23,Weingrill23, Vernekar25}, biasing our sample toward outer regions where the binary fraction is found to be lower \citep{HurleyAarsethShara07,Dalessandro11_M10,JiBregman15, Horn25_47Tuc}.

Compared to previous spectroscopic surveys \citep[e.g.][]{Sommariva09_M4, Lucatello15_BinaryGC, Wragg24_oCen, Horn25_47Tuc}, our analysis focuses on bright stars ($G_{\mathrm{RVS}} < 12$ mag), predominantly encompassing red giant and evolved stars in orbital periods typically shorter than \textit{Gaia}’s DR3 time coverage of $34$ months. 
While considering completeness corrections, we find in all GCs a binary fraction that, given the uncertainty, does not exceed $\sim 10\%$ in systems with orbital period $P \leq 10^4$ days.

In the case of M13, NGC 6397, and NGC 4833, the photometric study of \citet{Milone12} was the only available source used to estimate binary fractions until this work. Although the photometric method is sensitive to a broader range of orbital periods and inclinations, it is limited in mass ratio, and can give valuable results on systems with $q>0.5$. At the same time, it is focused mainly on stars along the main sequence. In contrast, our spectroscopic method, after applying completeness corrections, is primarily sensitive to evolved stars of short-period binaries $P<10^4$ days with a wider range of mass ratios. To ensure a consistent spatial comparison (Table~\ref{tab:GC_frac}), we selected in the case of M13 and NGC 6397 the stars located outside the half-mass radius defined by \cite{Milone12}, matching the annular region typically analysed by our \gaia sample. In the case of NGC 4833, the binary fraction reported was only available for an inner region of the cluster between the core and the half-mass radius, so we chose to cite this value instead.

A novel aspect of our study is the direct and homogeneous comparison of close binary fractions in the field and in GCs across a wide range of metallicities.
In the field, the close binary fraction shows a strong anti-correlation with [Fe/H], usually attributed to reduced disc-fragmentation efficiency at higher opacities \citep{Moe19, TokovininMoe20}. In GCs, and especially in their cores, by contrast, three-body captures, exchange encounters and tidal captures dominate both the formation and hardening of binaries, as predicted by Heggie’s law \citep{Heggie75,RoznerPerets24}. Compared with the clear trend in field stars, we find no strong correlation of binary fraction with [Fe/H] in GCs. 
Although metal-poor clusters are often assumed to be older and thus more dynamically processed, many metal-rich GCs that formed in situ are just as old as the metal-poor, accreted population \citep{BelokurovKravtsov24}. Our results therefore reinforce that cluster dynamics and evolutionary processes are the dominant regulators of binary statistics in GCs, whereas metallicity, so important in field binary demographics, plays a subdominant role within these dense stellar systems. This interpretation is consistent with previous photometric studies that found no significant correlation between cluster [Fe/H] and binary fraction \citep{Milone12}.


A related aspect which emerges from our analysis of \gaia data is that metal-poor stars exhibit noticeably higher RV jitter. 
One explanation is that weaker spectral lines in metal-poor stars produce larger RV uncertainties in the \gaia pipeline. Alternatively, the increased scatter could be astrophysical, driven by enhanced surface fluctuations or asteroseismic variability. Consistent patterns in APOGEE data suggest that this phenomenon is not unique to \textit{Gaia}. Future work incorporating an additional free parameter to model this offset should help clarify whether the increased jitter is predominantly due to measurement challenges or genuine stellar phenomena.

\section*{Acknowledgements}
We wish to thank Sebastian Kamann, the referee of this paper, for valuable comments which significantly improved our paper. We thank Alex Lobel, Paola Sartoretti, David Katz, and the entire CU6 team for their invaluable work on the \gaia RVS pipeline and for helpful discussions that significantly aided our analysis. We further thank Holger Baumgardt, Mark Gieles and Mor Rozner for their insightful comments on an early draft of this paper. We also thank Adrian Price-Whelan for his early feedback, which helped refine our analysis.
DB acknowledges the support of the Blavatnik family and the British Friends of the Hebrew University (BFHU) as part of the Blavatnik Cambridge Fellowship and Didier Queloz for his courteous hospitality.
VB acknowledges support from the Leverhulme Research Project Grant RPG-2021-205: ‘The Faint Universe Made Visible with Machine Learning’.

This work has made use of data from the European Space Agency (ESA) mission {\it Gaia} (\url{https://www.cosmos.esa.int/gaia}), processed by the {\it Gaia}
Data Processing and Analysis Consortium (DPAC,
\url{https://www.cosmos.esa.int/web/gaia/dpac/consortium}). Funding for the DPAC
has been provided by national institutions, in particular the institutions
participating in the {\it Gaia} Multilateral Agreement.

Funding for the Sloan Digital Sky Survey IV has been provided by the Alfred P. Sloan Foundation, the U.S. Department of Energy Office of Science, and the Participating Institutions. SDSS acknowledges support and resources from the Center for High-Performance Computing at the University of Utah. The SDSS web site is www.sdss4.org.
SDSS is managed by the Astrophysical Research Consortium for the Participating Institutions of the SDSS Collaboration including the Brazilian Participation Group, the Carnegie Institution for Science, Carnegie Mellon University, Center for Astrophysics | Harvard \& Smithsonian (CfA), the Chilean Participation Group, the French Participation Group, Instituto de Astrofísica de Canarias, The Johns Hopkins University, Kavli Institute for the Physics and Mathematics of the Universe (IPMU) / University of Tokyo, the Korean Participation Group, Lawrence Berkeley National Laboratory, Leibniz Institut für Astrophysik Potsdam (AIP), Max-Planck-Institut für Astronomie (MPIA Heidelberg), Max-Planck-Institut für Astrophysik (MPA Garching), Max-Planck-Institut für Extraterrestrische Physik (MPE), National Astronomical Observatories of China, New Mexico State University, New York University, University of Notre Dame, Observatório Nacional / MCTI, The Ohio State University, Pennsylvania State University, Shanghai Astronomical Observatory, United Kingdom Participation Group, Universidad Nacional Autónoma de México, University of Arizona, University of Colorado Boulder, University of Oxford, University of Portsmouth, University of Utah, University of Virginia, University of Washington, University of Wisconsin, Vanderbilt University, and Yale University.

This research also made use of TOPCAT \citep{Taylor05}, an interactive
graphical viewer and editor for tabular data.

This work made use of Astropy,
a community-developed core Python package and an
ecosystem of tools and resources for astronomy \citep{Astropy22}.

\section*{Data Availability}

The research presented in this article predominantly relies on data
that is publicly available and accessible online through the \gaia DR3
and APOGEE archives.



\bibliographystyle{mnras}
\bibliography{example} 




\appendix

\section{Binary-fraction modelling and completeness analysis}
\label{app:methods}

By exploiting the \gaia RV variability properties of a sample of stars, one can deconvolve the contributions of binaries and singles and estimate the overall close binary fraction. We choose to use the robust peak-to-peak RV amplitude $\mathrm{RV_{pp}}$, instead of the conventional standard deviation $\sigma_{\mathrm{RV}}$ originally used in \cite{Bashi24} to quantify the intrinsic RV variability of stars. This choice is motivated by the fact that \texttt{RV\_amplitude\_robust} is computed after removing outlier transits, ensuring that any anomalous measurements, typically affecting about $3\%$ of the bright \gaia RVS stars, do not skew the results. In contrast, $\sigma_{\mathrm{RV}}$ is calculated using all transit measurements, which can lead to an overestimation of the variability when even a few outliers are present. This provides a more reliable estimate of the intrinsic stellar RV variability, which is crucial for accurately distinguishing between single and binary star populations.

Given the dependence of the \gaia RV variability on apparent magnitude $G_{\rm RVS}$ \citep{Katz23, Bashi24, BashiTokovinin24}, we defined a density function as the sum of two Gaussian distributions: one for low-$\mathrm{RV_{pp}}/2$ (single stars) and one for high-$\mathrm{RV_{pp}}/2$ (binary stars). Each Gaussian was weighted by a binary fraction $F$. The other six free parameters in our model are then:  ii-iv) the single-star mean parameters ($a$, $b$, and $G_\text{0}$) that define $\mu_s(G_{\mathrm{RVS}})=a+e^{b(G_{\mathrm{RVS}}-G_\text{0})}$ to capture the dependence on source magnitude; (v) the additional variability parameter ($d$) that accounts for the extra RV variability in binaries; (vi-vii) and the standard deviations $\sigma_s$ and $\sigma_b$, which characterise the intrinsic spreads of the single-star and binary distributions, respectively. 

Parameter estimation was performed separately for each cluster’s data within a Bayesian framework using the \texttt{emcee} MCMC sampler \citep{Foreman-Mackey13}  with $40$ walkers and $10^4$ steps to estimate the parameter values and their uncertainties that maximise the sample likelihood. We list in Table~\ref{tab:priors} the priors used in our model.

\begin{table}
        \centering
        \caption{Prior distributions of the free parameters in our Bayesian Gaussian Mixture Model of single and binary stars.}
        \label{tab:priors}
        \begin{tabular}{lcc} 
                \hline \hline
                Parameter & Prior \\
                \hline
                $F$ & $\log \mathcal{U}(0.001,1)$ \\
                $a$ & $\mathcal{U}(-5,0.5)$& \\
                $b$ & $\log \mathcal{U}(10^{-3},1)$ \\
                $G_{\mathrm{min}}$ & $\mathcal{U}(0,15)$ \\
        $d$ & $\log \mathcal{U}(10^{-0.4},10^{1.6})$ \\
            $\sigma_s$ & $\log \mathcal{U}(0.001,10)$\\
            $\sigma_b$ & $\log \mathcal{U}(0.01,10)$\\
                \hline
        \end{tabular}
\end{table}

To quantify the completeness of our binary fraction estimates and find the fraction of missed binaries, we built a Monte-Carlo forward model that mimics every analysis step, from the observing cadence to the Bayesian mixture fit \footnote{We note that \citep{Bashi23} have conducted a similar analysis using both mock data and a simulated sample of single stars and binaries using the \gaia DR3 SB1 NSS sample \citep{NSS23}, and found the method is complete in estimating binary fractions of intermediate mass ratios systems, i.e., $q=0.3-0.9$ with orbital periods below $P \leq 10^3$ days.}. First, we draw $N_\star=5,000$ synthetic stars with magnitudes, $G_{\rm RVS}$, temperatures, gravities and metallicities sampled from smooth kernel estimates of the observed distributions for each cluster. Stellar masses are assigned with the empirical \cite{Torres10} relation.
For every star, we pick the number of RVS visits from a Poisson distribution with expectation value of 10, and enforce a minimum of eight transits matching the actual DR3 cuts we used. These visits are placed at random within a 34-month window, reproducing the DR3 time baseline.
Each star is flagged as binary with probability $F_{\rm true}$. If so, we draw orbital parameters as follows:
orbital period is drawn from a log-uniform distribution, $P\sim \log \mathcal{U}(1, P_{\rm max})$, where $P_{\rm max}$ is the maximal period of the binary under test; mass ratio  is drawn from a uniform
distribution, $q\sim \mathcal{U},(0.01,0.9)$ where the upper limit of $0.9$ is aimed to avoid cases of equal-mass stars with opposite RV variations making it
more difficult to detect the system spectroscopically using \gaia RVS; binary inclination is drawn from a uniform distribution using $\cos i \sim \mathcal{U}(0,1)$, and argument of periastron is drawn uniformly using, $\omega \sim \mathcal{U}(0,2\pi)$. Eccentricity, $e$ is generated in two steps: binaries with $P\le P_{\rm cut}=5$ days are assumed circular to reflect efficient tidal damping \citep{Bashi23}. For longer periods, we sample $e$ from a Rayleigh distribution with scale $\sigma=0.3$, consistent with field binaries of similar spectral type \citep{Raghavan10}. 
From these, we compute the RV semi-amplitude $K$, and generate instantaneous velocities at random epochs using a Kepler solver.
Each synthetic RV is perturbed with Gaussian noise whose standard deviation follows the empirical RVS error model
$\sigma_{\rm RV}=10^{\,\log a + b\,(G_{\rm RVS}-G_0)}$,
with $(\log a,b,G_0,\sigma_{\rm s})$ fixed to the median posterior values obtained for the real data.
For every star we compute our working observable, $\mathrm{RV_{pp}}/2$, and re-run the full MCMC mixture-model pipeline. The recovered median binary fraction, $F_{\rm est}$, is compared to the injected $F_{\rm true}$.

\begin{figure}
   \centering
   \includegraphics[width=0.5\textwidth] {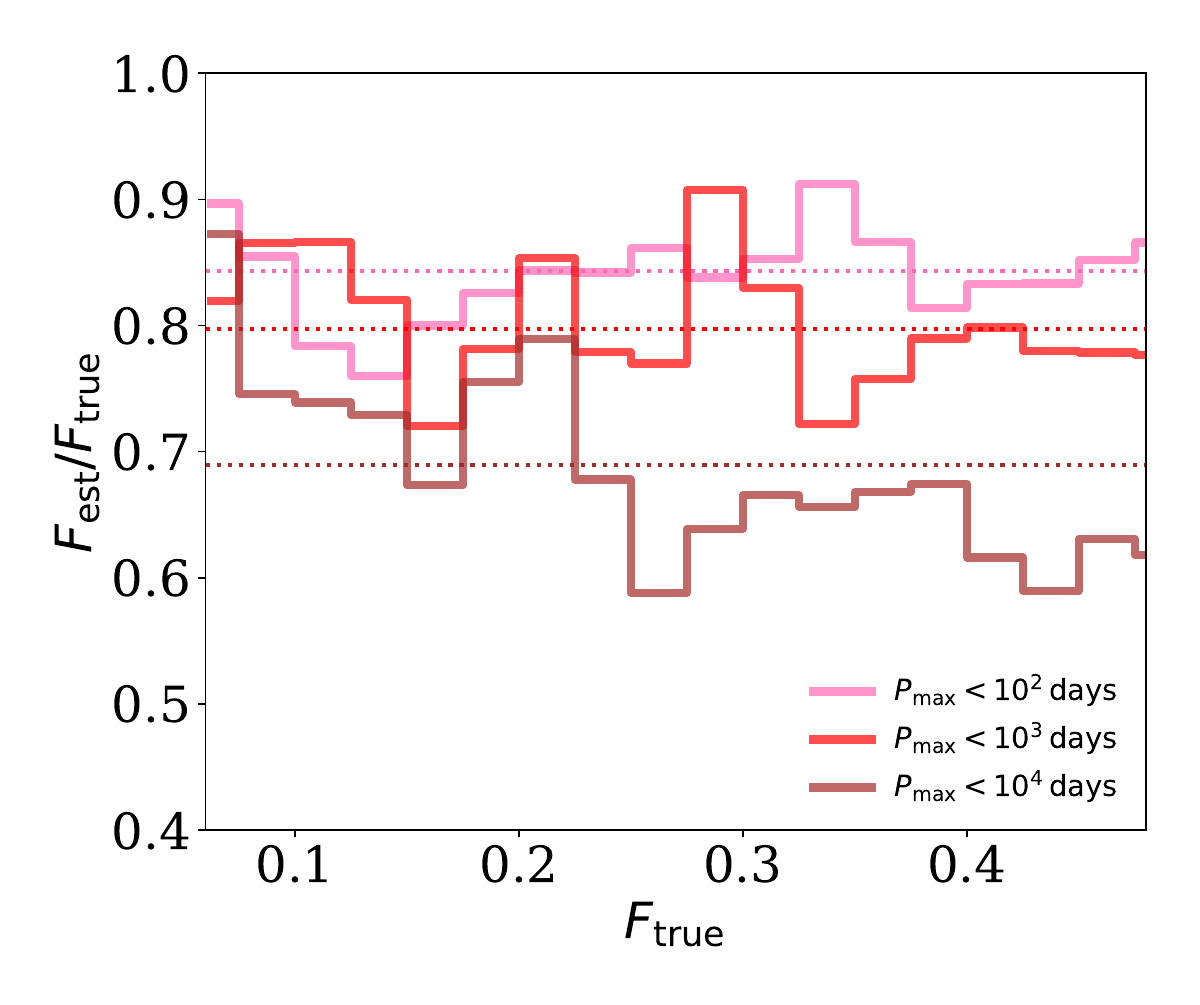}
   \caption{Recovery efficiency of the \textit{Gaia}-RVS binary-fraction estimator.
Ratio of the recovered to injected binary fraction, $F_{\mathrm{est}}/F_{\mathrm{true}}$, as a function of the input fraction $F_{\mathrm{true}}$ derived from 500-star Monte-Carlo catalogues that reproduce the DR3 cadence, noise model and magnitude distribution (see text for further details). Curves are colour-coded by the maximum orbital period $P_{\mathrm{max}}$ allowed in the mock population.} 
        \label{fig:binary_fraction_ratio_q0p01_0p9}%
    \end{figure}

Repeating the experiment for a range of input fractions $F_{\rm true}$ and three upper period values ($P_{\rm max}<10^2$ days; $P_{\rm max}<10^3$ days; $P_{\rm max}<10^4$ days) we show in Fig.~\ref{fig:binary_fraction_ratio_q0p01_0p9} the ratio of the recovered to injected binary fraction, $F_{\mathrm{est}}/F_{\mathrm{true}}$, as a function of the input fraction $F_{\mathrm{true}}$. As expected, as we go to longer orbital periods, the relative fraction decreases on average as we miss more long-period binaries given their lower semi-amplitude signal and the lack of full coverage of the full orbital motion given the \gaia DR3 phase window.

The mean recovery factors $\varepsilon$ in each maximum orbital period distribution are then the reverse of the $F_{\mathrm{est}}/F_{\mathrm{true}}$ ratio, which are: $\varepsilon_{P_{100}} = 1.18$;
$\varepsilon_{P_{1000}} = 1.25$;
$\varepsilon_{P_{10000}} = 1.45$.
Thus, the completeness-corrected intrinsic close binary fraction is simply $\bar{F}=\varepsilon F$, where $F$ is the fraction returned by our MCMC mixture model.

\bsp	
\label{lastpage}
\end{document}